\documentclass[amsmath,amssymb,prb,twocolumn,amssymb,eqsecnum]
{revtex4}
\usepackage{graphicx}
\usepackage{verbatim}
\usepackage{color}

\DeclareMathOperator{\im}{Im}
\DeclareMathOperator{\re}{Re}

\DeclareMathOperator{\var}{Var}
\DeclareMathOperator{\artanh}{artanh}

\begin{document}
\title{Frequency-dependent admittance of a short superconducting weak link} 
\author{F.~Kos, S.~E.~Nigg, and L.~I.~Glazman}
\affiliation{Department of Physics, Yale University, New Haven, CT 06520, USA}
\date{\today}
\begin{abstract}
We consider the linear and non-linear electromagnetic responses of a nanowire connecting two bulk superconductors. Andreev states appearing at a finite phase bias substantially affect the finite-frequency admittance of such
a wire junction. Electron transitions involving Andreev levels are easily saturated, leading to the nonlinear effects in photon absorption for the sub-gap photon energies. We evaluate the complex admittance analytically at arbitrary frequency and arbitrary, possibly non-equilibrium, occupation of Andreev levels. Special care is given to the limits of a single-channel contact and a disordered metallic
weak link. We also evaluate the quasi-static fluctuations of
admittance induced by fluctuations of the occupation factors of
Andreev levels. In view of possible qubit applications, we compare
properties of a weak link with those of a tunnel Josephson junction. Compared to the latter, a weak link has smaller low-frequency dissipation. However, because of the deeper Andreev levels, the low-temperature quasi-static fluctuations of the inductance of a weak link are exponentially larger than of a tunnel junction. These fluctuations limit the applicability of nanowire junctions in superconducting qubits. 
\end{abstract}
\maketitle

The search for longer coherence times of superconducting qubits
brought the study of finite-frequency electromagnetic properties of
mesoscopic superconductors to the forefront of experimental research
~\cite{klapwijk, martinis1,siddiqi,schoelkopf1,mcdermott,clarke}. The
majority of experiments until recently was performed on
structures using Josephson junctions as ``weak'' superconductors, and substantial progress in recognizing the coherence-limiting
mechanisms was achieved. One may view a number of mechanisms causing
energy or phase relaxation as extrinsic ones. These  involve,
 {e.g.}, imperfections in the tunnel barriers comprising junctions~\cite{martinis4}, charge trapping~\cite{delsing}, and interaction with stray photons~\cite{schoelkopf2,schoelkopf3}. Along with them, there are intrinsic mechanisms associated with the kinetics of quasiparticles in the superconductors~\cite{lutchyn,catelani1,martinis2}. These mechanisms provide fundamental limitations to the coherence. The majority of effects of quasiparticles on the finite-frequency properties of Josephson junctions can be derived~\cite{martinis3,catelani1} from the electromagnetic admittance of the junction $Y(\omega)$. This property was extensively studied theoretically, starting from the seminal phenomenological paper of Josephson ~\cite{josephson} and microscopic evaluation~\cite{werthamer,LO} based on the BCS theory. 

The use of weak superconducting links instead of Josephson junctions
in qubits was proposed recently as a way to avoid extrinsic
decoherence mechanisms (such as imperfections of the tunnel
barriers)~\cite{mooij}. An apparent observation of a coherent phase
slip in a conducting weak link~\cite{astafiev} may be viewed as an
incipient experimental step in that direction. That makes the question
about the intrinsic mechanisms of decoherence in weak links
important. Like with Josephson junction devices~\cite{catelani1}, this question is
directly related to the finite-frequency admittance of a weak
link. Surprisingly, this property was given relatively little
attention to. The admittance of a short SNS contact was investigated,
mostly numerically, in the recent papers~\cite{cuevas,
  heikkila}. Some qualitative aspects of the AC response of a single-channel point contact can be extracted from two other papers devoted to the theory of enhancement of supercurrent by microwave radiation~\cite{bergeret1,bergeret2}. 

Here we perform a fully-analytical evaluation of the admittance of a
weak link connecting two bulk superconductors, valid at arbitrary
frequency $\omega$, quasiparticle distribution function, and
normal-state conductance of the link. Compared to the Josephson
junction case, the dissipative part of the weak link admittance
exhibits a number of new thresholds in its frequency dependence,
associated with the presence of Andreev levels. The complex admittance
close to these new threshold frequencies is sensitive to the
occupation of the discrete Andreev states. Fluctuations of the
equilibrium or non-equilibrium occupation factors result in
fluctuations of the admittance. We analyze the average values and
fluctuations of the linear electromagnetic response, giving special
attention to the practically important limits of a single-channel
contact~\cite{urbina} and a disordered metallic wire~\cite{siddiqi,siddiqi2}.

The discrete nature of Andreev states is responsible for a low threshold for the nonlinear absorption. In the nonlinear regime, we find a suppression of the absorption coefficient in a disordered metallic link at radiation frequency  $\omega\leq2\Delta/3$, while at higher $\omega$ dissipation power depends non-linearly on the radiation intensity (here $\Delta$ is the BCS gap in the leads).

The paper is organized as follows: the model used in the derivation of the admittance of a point contact with an arbitrary transmission coefficient is formulated in Section~\ref{pcH}. The linear response theory for the AC perturbation of the point contact is developed in Section~\ref{linres}. In Section~\ref{singlech} we discuss the results for the admittance of the point contact at zero temperature and no quasiparticles present. In Section~\ref{quasip} we study the changes in the admittance caused by the arbitrary distribution of quasiparticles in the junction. These results are used in Section~\ref{weaklink} to find the admittance of a disordered weak link. The fluctuations of the admittance are analyzed in Section~\ref{fluctuation}, both for the case of point contact and of a weak link. In Section~\ref{nonlinabs} we consider the absorption rate in a non-linear regime for the radiation frequencies close to the Andreev level resonance. We conclude with the final remarks in Section~\ref{conclusion}.

\section{Point contact Hamiltonian}
\label{pcH}
We start by considering a point contact between two leads. It can be
described by the tunnel Hamiltonian
\begin{equation}
\label{H}
\hat H = \hat H_L + \hat H_R + \hat H_T\,,
\end{equation}
where $\hat H_{L(R)}$ are the BCS Hamiltonians of the left (right) leads:
\begin{eqnarray}
&\hat H_{L} =\sum_k \xi _k c^\dagger _k c_k + \Delta_L \sum_k c^\dagger _k c^\dagger _{\bar k} + \Delta_L^* \sum_k c_{\bar k} c_k\,,
\label{HL}\\
&\hat H_R = \sum_p \xi _p c^\dagger _p c_p + \Delta_R \sum_p c^\dagger _p c^\dagger _{\bar p} + \Delta_R^* \sum_p c_{\bar p} c_p\,,
\label{HR}
\end{eqnarray}
and
\begin{equation}
\label{HT}
\!\!\hat H_T = w \!\sum_{kp} \left(c^\dagger_k c_p \!+\! c^\dagger_p c_k\right)\!+ w\!\sum_{k_1k_2} c^\dagger_{k_1} c_{k_2}\!+w\! \sum_{p_1p_2} c^\dagger_{p_1} c_{p_2}
\end{equation}
is the tunneling term. Here $c_{k(p)}$ and $c_{{\bar k},({\bar p})}$ are electron operators in the left (right) lead corresponding to states $k (p)$ and its time-reversed pairs ${\bar k}({\bar p})$, and $\Delta_{L(R)} = \Delta e^{i\phi_{L(R)}}$ are the BCS gap functions. 

The tunneling amplitude $w$ is assumed to be
momentum-independent near the Fermi level. It is related to
the transmission coefficient, $\tau = (2\pi\nu_0w)^2/[1
+ (2\pi \nu_0 w)^2]$, where $\nu_0$ is normal-state density of
states. The conductance of the junction $G$ in the normal state is
proportional to $\tau$ (hereinafter we set $\hbar=1$). A point contact between superconducting leads hosts a single Andreev level with energy $E_A(\tau, \phi)$ depending on $G$:
\begin{equation}
G= e^2\tau/\pi \, , \quad E_A (\tau, \phi) = \Delta (1-\tau \sin^2\phi/2)^{1/2}\,; \label{GEA}
\end{equation}
here the phase difference between the leads order parameters, $\phi=\phi_R - \phi_L$, is assumed to be time-independent.

\section{Linear response to AC perturbation}
\label{linres}
We may account for an applied small, time-dependent voltage $V(t)$ by
modifying $\phi_L\to\phi_L+2\phi_1(t)$ in Eq.~(\ref{HL}), with
$\dot{\phi}_1=eV(t)$, and adding the term $-eV(t)\hat{N}_L$ to Eq.~(\ref{H}):
\begin{equation}
\mathcal{H} =\hat H(t) - eV \hat N_L\,,\quad \hat{N}_L=\sum_kc_k^\dagger c_k^{\phantom{\dagger}}\,.
\label{calH}
\end{equation}
We want to find the current $\langle \hat{I}\rangle$,
\begin{equation}
\label{hatI}
\hat I = e \dot N_L = -i e w \sum_{kp} \left[ c^\dagger_{k} d_{p} - d^\dagger_{p}c_{k}\right]\,,
\end{equation}
induced by an applied voltage to linear order in $V$ and at arbitrary transmission $\tau$. The validity of linear response in $V$ requires at least the smallness of the perturbation to the dynamics of the system, $|\phi_1|=|eV/\omega|\ll 1$, where $\omega$ is the frequency of perturbation. Further limitations on the parameters, which may come from the effect of $V$ on occupation factors, will be discussed later.

It is convenient to do the gauge transformation $c_k \to c_k
e^{i\phi_1}$ before performing the perturbation theory. This moves the
$\phi_1$--dependence to the tunneling terms. Using the Kubo formula for linear response, we get
\begin{equation}
\langle \hat I (t) \rangle = I_J + \int_{-\infty}^\infty dt' \chi (t-t') \phi_1(t') \, .
\end{equation}
Here $I_J$ is the Josephson current which is present even without applied voltage:
\begin{equation}
I_J = e w \im \Big\langle \sum_{kp} c^\dagger_k d_p \Big\rangle \label{IJ1}\, .
\end{equation}
The response function $\chi(t)$ is given by
\begin{align}
&\chi(t) =   i e w^2 \theta(t)  \sum_{k_1p_1}\sum_{k_2p_2} \Big\langle \big[  c^\dagger_{k_1}(t) d_{p_1}(t) - d^\dagger_{p_1}(t)c_{k_1}(t),   \notag\\ 
& c^\dagger_{k_2}(0) d_{p_2}(0) - d^\dagger_{p_2}(0)c_{k_2}(0) \big]\Big\rangle  -e w \delta(t) \re \Big\langle \sum_{kp} c^\dagger_{k} d_{p} \Big\rangle \, .  \label{chi2}
\end{align}
Averages $\langle \dots\rangle$ are taken over the Gibbs ensemble of the
original Hamiltonian $H$. We can use Wick's theorem to evaluate
averages. They can be expressed in terms of Green's functions of the
unperturbed system. Green's functions satisfy a system of
linear integral equations, but the corresponding kernels are separable due to the form of the tunneling term Eq.~(\ref{HT}). Therefore, that system reduces
to a system of algebraic equations which can be solved
exactly. This response function is related to the admittance in frequency domain:
\begin{equation}
Y(\omega) = i\frac{e}{\omega}\chi(\omega)\,. \label{Yomega}
\end{equation}
It is convenient to split the admittance into a sum
\begin{equation}
Y=\sum_{i=1}^5Y_i+ \frac{i}{\omega L_J} \, , \label{Ysum}
\end{equation}
each term of which has a clear physical origin. The purely inductive
term $L_J$ comes from the $\omega\to 0$ response of the condensate,
\begin{equation}
\frac{1}{L_J}=2e\frac{\partial I_J}{\partial\phi} \, .
\label{LJ}
\end{equation}
The other five parts of Eq.~(\ref{Ysum}) originate from the quasiparticles transitions. To better understand the structure of these parts, recall that $\re Y$ (dissipative part of admittance) is related to the linear absorption rate $W$ of the radiation by:
\begin{equation}
W = \frac{\phi_1^2}{2e^2}| \omega | \re Y(\omega)\,. \label{absorptionrate}
\end{equation}
The elementary processes leading to the absorption are depicted in Fig.~1. The $Y_1(\omega)$ term corresponds to a process in which two quasiparticles are created in the band, leading to the energy threshold $2\Delta$.  The contribution $Y_2(\omega)$ in (\ref{Ysum}) comes from creating one
quasiparticle in the bound state and one in the band, the
corresponding threshold energy is $\Delta + E_A$. Creation of a pair
of quasiparticles in the bound state~\cite{rodero}, which costs energy
$2E_A$, leads to the term $Y_3(\omega)$. In addition to these three contributions which exist even in the absence of quasiparticles, there are two more associated with the promotion of an existing quasiparticle to a higher energy in the absorption process: $Y_4$ is the intra-band contribution, and $Y_5$ corresponds to an ionization of an occupied Andreev level.

\section{Admittance of a single-channel junction at $T=0$ (equilibrium state with no quasiparticles)} 
\label{singlech}
Evaluating averages in Eq.~(\ref{IJ1}) for system at zero temperature we get for the Josephson current:
\begin{equation}
I_J^{(0)}=  \pi G  \frac{\Delta^2\sin\phi}{2e E_A(\phi)} \, , \label{IJ0}
\end{equation}
in agreement with the result one obtains from Eq.~(\ref{GEA}) by differentiating energy over $\phi$, i.e. $I_J^{(0)} = -2e \partial E_A/\partial \phi$. Using the above expression for $I_J^{(0)}$ and Eq.~(\ref{LJ}) we find the inductive term at zero temperature:
\begin{align}
\frac{1}{L_J^{(0)}}  = \pi G \Delta\frac{\cos \phi +\frac{\pi G}{e^2} \sin^4\frac{\phi}{2}}{\left(1 - \frac{\pi G}{e^2}\sin^2 \frac{\phi}{2} \right)^{3/2}} \label{LJ0} \, .
\end{align}

\begin{figure}
\begin{center}
\includegraphics[scale=0.3]{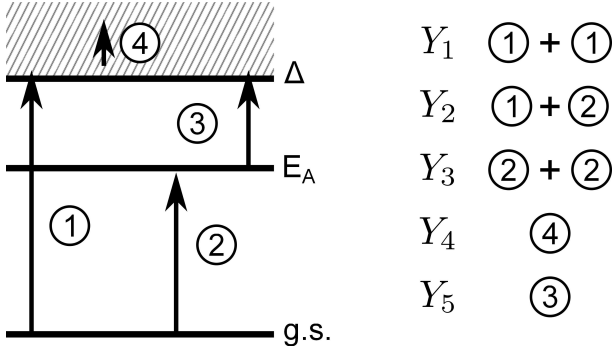}
\caption{Schematic representation of contributions to the admittance, Eqs.~(\ref{reY1})-(\ref{reY3}),(\ref{reY4}) and (\ref{reY5}). Horizontal bars are energy levels and arrows depict possible excitations. The picture shows that $Y_4 = Y_5 = 0$ in the absence of quasiparticles. Since $\re Y (\omega)$ is proportional to the absorption rate, we see that $\re Y_{1,2,3,5}$ will have threshold frequencies of $2\Delta$,$\Delta+E_A$, $2 E_A$, $\Delta- E_A$, respectively. }\label{fig1}
\end{center}
\end{figure}

Evaluating averages in Eq.~(\ref{chi2}) we find the contributions $Y_i^{(0)}$ to the admittance. With no quasiparticles present, there can be no processes of type 3 or 4 in Figure~\ref{fig1}. Therefore, $Y_4^{(0)} (\omega)= Y_5^{(0)}(\omega)=0$. The contribution $Y_1^{(0)}(\omega)$  comes from the creation of pairs of quasiparticles in the band (two excitations of type 1 in Figure~\ref{fig1}), and its real part is given by:
\begin{align}
\frac{\re Y_1^{(0)} (\omega)}{G} = &\frac{\theta(\omega- 2\Delta)}{\omega}  \int_\Delta^{\omega- \Delta} d\epsilon \rho (\epsilon) \rho (\omega - \epsilon) |z_{\epsilon, \omega}|^2\, , \label{reY1}
\end{align}
where $\rho(\omega)$ is the density of states in the continuum normalized to normal-state density of states $\nu_0$:
\begin{align}
\rho(\epsilon) =  \frac{\epsilon \sqrt{\epsilon^2 - \Delta^2}}{\epsilon^2 - E_A^2}\, ,
\label{rho}
\end{align}
and the matrix element $z$ is given by:
\begin{align}
|z_{\epsilon,\omega}|^2 = 1 -\frac{\Delta^2\cos\phi+ \Delta^2 -E_A^2}{\epsilon( \omega - \epsilon)}\,.
\label{z-ep-omega}
\end{align}
We assume $\omega>0$ throughout this section. The result for negative frequencies can be found using the fact that $\re Y^{(0)} (\omega)$ is an even function. The theta function in Eq.~(\ref{reY1}) shows that there can be no creation of pairs in the continuum for frequencies less than $2\Delta$.  

The $Y_2^{(0)}(\omega)$ term  comes from processes in which one quasiparticle is created in the band and another one in the Andreev level. These processes are represented by one arrow of type 1 and one of type 2 in the Figure~\ref{fig1}. The real part of $Y_2^{(0)}(\omega)$ is given by:
\begin{align}
&\frac{\re Y_2^{(0)} (\omega)}{G} \!= \! \pi \theta(\omega\!\!-\!\! E_A\!\! -\! \! \Delta) \frac{\sqrt{\!\Delta^2\! - \! E_A^2}}{\omega}\rho(\omega \!-\! E_A) |z_{E_A, \omega}|^2 \,, \label{reY2}
\end{align}
and it vanishes for $\omega < \Delta + E_A$, as for these frequencies the processes ``1+2" are energetically not allowed.

Finally, there are processes in which two quasiparticles on Andreev level are created. Those are represented by two excitations of type 2 in Figure~\ref{fig1}. In this case, the frequency must be equal to $2E_A$. The $Y_3^{(0)}(\omega)$ term comes from such processes and its real part is given by:
\begin{align}
&\frac{\re Y_3^{(0)}(\omega)}{G}=\pi^2 \frac{(\Delta^2 \!-\! E_A^2)(E_A^2 \!-\! \Delta^2 \cos^2\frac{\phi}{2})}{2 E_A^3}\delta(\omega \!-\! 2E_A) \,. \label{reY3}
\end{align}
Note that the RHS of Eqs.~(\ref{reY1})-(\ref{reY3}) depend on $G$ and $\phi$ through $E_A$, see Eq.~(\ref{GEA}).

The admittance exhibits non-analytical behavior at threshold frequencies $\omega = 2E_A$, $\Delta + E_A$ and $2\Delta$. For $\omega \approx 2\Delta$ we have $\re Y_1^{(0)} \propto (\omega-2\Delta)^2 \theta (\omega-2\Delta)$ according to (\ref{reY1}). Similarly, for frequencies $\omega \approx \Delta + E_A$ we get $\re Y_{2}^{(0)} \propto \sqrt{\omega - (\Delta + E_A)}\theta(\omega - (\Delta + E_A))$ from Eq.~(\ref{reY2}). 

The imaginary parts of $Y_i(\omega)$'s can be obtained from their real parts using Kramers-Kronig relations since $Y(\omega)$ is analytic in the upper half of the complex $\omega$-plane. The complete expression for $\im Y(\omega)$ is given in Appendix A. At threshold frequencies $\im Y(\omega)$ exhibits non-analytical behavior which parallels threshold behavior of $\re Y (\omega)$. At $\omega \approx 2 \Delta$ the non-analytical contribution behaves as $\im Y_1^{(0)} \propto (\omega-2\Delta)^2 \ln |\omega - 2 \Delta|$ and at $\omega \approx \Delta+ E_A$ it behaves as $\im Y_2^{(0)} \propto \sqrt{(\Delta + E_A) - \omega}\theta( (\Delta + E_A) - \omega ))$. The coefficients omitted from the asymptotes of $\re Y^{(0)}(\omega)$ and $\im Y^{(0)} (\omega)$ equal each other, confirming that the complex function $Y^{(0)}(\omega)$ is analytical.

\section{Admittance of a single-channel junction in the presence of quasiparticles}
\label{quasip}
The admittance changes once there are quasiparticles present. Each term in Eq.~(\ref{Ysum}) acquires  an additional factor depending on the quasiparticle occupation numbers. We introduce occupation factors $p_0$, $p_\uparrow$, $p_\downarrow$ and $p_2$ denoting probabilities of having zero, one or two quasiparticles in the bound state; $p_0+p_\uparrow + p_\downarrow+p_2=1$. The inductance in Eq.~(\ref{Ysum}) then becomes:
\begin{align}
&\frac{1}{L_J} = \frac{1}{L_J^{(0)}} [p_0-p_2] \,.  \label{LJ1}
\end{align}
The $Y_1$ term acquires a factor depending on the occupation factors $f(\epsilon)$ of the continuum states, 
\begin{align}
\frac{\re Y_1^{(0)} (\omega)}{G} = &\frac{\theta(\omega- 2\Delta)}{\omega}  \int_\Delta^{\omega- \Delta} d\epsilon \rho (\epsilon) \rho (\omega - \epsilon) |z_{\epsilon, \omega}|^2 \notag\\
&\times [1 - f(\epsilon) - f(\omega - \epsilon)]\,. \label{factor1}
\end{align}
This expression is different from Eq.~(\ref{reY1}) by a factor equal to the difference of probabilities for having the initial and the final state occupied in the transition from the ground state to the band, see Fig.~\ref{fig1}. Similarly, the $\re Y_2$ term is given by
\begin{equation}
\re Y_2(\omega) = \re Y_2^{0}(\omega) (p_0 + \frac{p_\uparrow+p_\downarrow}{2}-f(\omega-E_A)) \label{factor2}\, ,
\end{equation}
and $Y_3$ term by
\begin{equation}
Y_3(\omega) = Y_3^{0}(\omega) (p_0 -p_2) \,.
\end{equation}

At non-zero occupancies, there are two additional contributions to absorption, $Y_4(\omega)$ and $Y_5(\omega)$. The former one comes from the band-to-band transitions, represented by the arrow 4 in Figure~\ref{fig1}. Its real part is given by:
\begin{align}
\frac{\re Y_4(\omega)}{G} =&  \frac{2}{\omega} \int_\Delta^\infty d\epsilon \rho(\epsilon) \rho(\omega \!+ \! \epsilon)|z_{-\epsilon,\omega}|^2  [ f(\epsilon)\!-\!f(\omega\!+\!\epsilon) ] \,.\label{reY4}
\end{align}
 The other term, $Y_5(\omega)$, is generated by the Andreev level-to-band transitions~\cite{bergeret1}. These transitions are represented by the arrow 3 in Figure~\ref{fig1}. We can write it in the form resembling that of $Y_2(\omega)$:
\begin{align}
\re Y_5(\omega) = \re Y_5^{(0)}(\omega)  [p_\uparrow+p_\downarrow+2p_2 - 2 f(\omega \!+\! E_A) ]\, , \label{reY5n}
\end{align}
where the real part of $Y_5^{(0)}(\omega)$ is given by:
\begin{align}
&\frac{\re Y_5^{(0)}\! (\omega)}{G} \!= \! \pi \theta(\omega\!\!+\!\! E_A\!\! -\! \! \Delta) \frac{\sqrt{\!\Delta^2\!\! -\! \! E_A^2}}{2 \omega}\rho(\omega \!+\! E_A) |z_{-\!E_A, \omega}|^2 \,, \label{reY5}
\end{align}
The occupation factors $p$ and $f$ in all of the above expressions may, but need not to be the equilibrium ones. The density of states $\rho(\epsilon)$ and matrix element $z_{\epsilon,\omega}$ in Eqs.~(\ref{factor1}) and (\ref{reY4})-(\ref{reY5}) are defined in Eqs.~(\ref{rho}) and (\ref{z-ep-omega}).

Note that  the transitions involving the Andreev level vanish at $E_A \to \Delta$. This is achieved if either $G=0$ or $\phi = 0$. Expanding to the lowest order in $G$ for $G\ll e^2/\pi$, the contributions $Y_{1,4}$ are linear in $G$ and they reduce to the familiar perturbative result for the admittance of a Josephson junction~\cite{barone}. The other contributions are higher order in $G$, with $Y_{2,5} \propto G^{3/2}$, and $\re Y_3 \propto G^2 $.

In the limit $\phi \to 0$, the terms involving transitions to Andreev level vanish as $\re Y_{2,5} \propto |\phi|$, and $\re Y_3 \propto \phi^2$. Therefore, at $\phi =0$ only terms contributing to $Y$ are again $Y_1$ and $Y_4$. In that case, the expression for $Y$ coincides with the one found perturbatively~\cite{barone}  in the limit of weak tunneling $G\ll e^2/\pi$. Thus, at small phase bias $\phi$ we don't expect much difference from the simple Josephson junction.

Now we analyze the behavior of admittance in the limits of low frequencies  and low temperatures, as these are the conditions often encountered in the application of the superconducting junctions. At frequencies below the threshold for Andreev level ionization, $\omega < \Delta - E_A$, and away from the bound pair creation resonance, $\omega\neq 2E_A$, the only contribution to the dissipative part of admittance comes from  Eq.~(\ref{reY4}). We assume the quasiparticle occupation factors are distributed according to Boltzmann distribution $f(\epsilon)=Ae^{-\epsilon/T}$.  At low temperatures $T \ll \Delta-E_A$, where $\Delta - E_A$ is characteristic scale for the energy dependence of the density of states above the gap, the dominant contribution to $\re Y_4(\omega)$ comes from the transitions between the states near the bottom of the band. In that limit, we get for the asymptotic form of $\re Y_4(\omega)$:
\begin{align}
\frac{\re Y_4 (\omega)}{G} \approx &\sqrt{\frac{2}{\pi}}x_{qp}\frac{1+\cos\phi + \frac{\pi G}{e^2}\sin^2\frac{\phi}{2}}{(\frac{\pi G}{e^2})^2 \sin^4 \frac{\phi}{2}} \notag\\ 
&(1-e^{-\omega/T}) \sqrt{\frac{T}{\Delta}} e^{\omega/4T}K_1(\frac{\omega}{4T}) \, , \label{reY4as}
\end{align}
where $x_{qp}$ is the density of quasiparticles, $n_{qp}$, in the bulk normalized to the ``Cooper pair density'', $x_{qp}=n_{qp}/\nu_0 \Delta$, and $K_1(x)$ is the modified Bessel function of the second kind. Note that at small frequencies,  $\omega \ll T$, it follows from Eq.~(\ref{reY4as}) that $\re Y_4 (\omega)$ is frequency-independent and proportional to $ \sqrt{T/\Delta} $.

In the limit $G \ll e^2/\pi$, the Andreev level is shallow, $\Delta - E_A \ll \Delta$. If  now $T \gg \Delta - E_A$, the main contribution to $Y_4$ comes from transitions involving states far above the gap where the density of states is described by the usual BCS result. In this limit, Eq.~(\ref{reY4}) is reduced to the known result~\cite{catelani1} for a Josephson junction,
\begin{equation}
\frac{\re Y(\omega)}{G} \approx \frac{1}{2\sqrt{2}} \frac{n_{qp}}{\nu_0\Delta} (1+\cos\phi) \left( \frac{\Delta}{T} \right)^{3/2} \ln\frac{4 T}{\omega} \,.\label{reYtj}
\end{equation}
This limit is the opposite to the one of Eq.~(\ref{reY4as}). The two asymptotes match each other at $T\sim \Delta- E_A$  up to the logarithmic factor.

It is interesting to compare the dissipation in a large-area Josephson
junction of $G\lesssim e^2/\pi$ with the dissipation in a
single-channel weak link of the same $G$. The weak-link quasiparticle
density of states in the continuum, see Eq.~(\ref{rho}), is suppressed
compared to the singular tunneling density of states in a Josephson
junction. As a result, at frequencies and temperatures $\omega,
T\lesssim \Delta-E_A$  a weak link is less dissipative than a
Josephson junction with small-transparency but large-area tunnel
barrier of the same $G$. Using Eqs. (\ref{reY4as}) and (\ref{reYtj})
we find, e.g., that the dissipation is smaller by a factor
$T^2/\Delta^2$ in the case of a weak link.

\section{Disordered weak link}
\label{weaklink}
For a multi-channel junction one needs to sum the contributions to the admittance from each channel. We consider the case of a disordered weak link for which we can assume the transmission coefficients are continuously distributed according to Dorokhov distribution~\cite{dorokhov} $\rho(\tau) = \pi G/2 e^2 \tau \sqrt{1-\tau}$. The admittance is then given by:
\begin{equation}
\bar Y (\omega) = \int_0^1 d\tau \rho (\tau) Y (\omega,\tau) \label{Ybar} \,.
\end{equation}
We can write $\bar Y (\omega)$  as a sum of five terms, in the same
way we did it for the single-channel junction in
Eq~(\ref{Ysum}). Transitions between Andreev levels are ignored, which
is justified in the limit of short junction with $\Delta/E_T \to
0$, where $E_T$ is Thouless energy~\cite{levchenko}.

The Josephson current of the disordered weak link can be found using the same averaging procedure as in Eq.~(\ref{Ybar}). In the absence of quasiparticles it is given by:
\begin{align}
\bar I _J^{(0)} = \frac{\pi G \Delta}{e}\cos\frac{\phi}{2}\artanh{\sin\frac{\phi}{2} }\,. \label{IJs}
\end{align}
Similarly, averaging $1/L_J^{(0)}$ from  Eq.~(\ref{LJ0}) we get:
\begin{align}
\frac{1}{\bar L_J^{(0)}} =\pi G\Delta \left[ 1- \sin\frac{\phi}{2}\artanh{\sin\frac{\phi}{2} }\right]\,. \label{LJs}
\end{align}
Evaluating the integral in Eq.~(\ref{Ybar}) results in expressions for
$\bar Y_i$.  If there are no quasiparticles present, the only
non-vanishing terms are $\bar Y_{1,2,3}$. Their real parts exhibit
threshold behavior at frequencies $2\Delta$, $ \Delta +
\Delta|\cos\frac{\phi}{2}|$ and $2 \Delta
|\cos\frac{\phi}{2}|$. The latter two thresholds correspond to the fully transmitting channel which has the lowest possible Andreev level energy at given phase $\phi$. The term corresponding to the creation of a pair of quasiparticles in the continuum is given by:
\begin{widetext}
\begin{align}
\frac{\re \bar Y_1^{(0)}(\omega)}{G} =\theta(\omega - 2\Delta)\text{P.V.} \!\!\!\! \int\limits_{\Delta}^{\omega- \Delta}\!\! d\omega_1  \frac{\omega_1\omega - \Delta^2 (1 + \cos\phi)  }{ \Delta^2 \omega^2 | \sin\frac{\phi}{2}|( \omega- 2\omega_1)  } \frac{ \sqrt{\omega_1^2 \!-\! \Delta^2\!} \sqrt{\!(\omega \!-\! \omega_1\!)^2 \!-\! \Delta^2\!} }{\sqrt{\omega_1^2 / \Delta^2 - \cos^2\frac{\phi}{2}}}  \ln \frac{\sqrt{\omega_1^2/\Delta^2 - \cos^2\frac{\phi}{2}} + |\sin \frac{\phi}{2}|}{ \sqrt{\omega_1^2/\Delta^2 - \cos^2\frac{\phi}{2}} - |\sin \frac{\phi}{2}| } \,. \label{reYbar1}
\end{align}
\end{widetext}
It has the threshold frequency $2\Delta$, just like the single channel admittance, Eq. (\ref{reY1}). At frequencies higher than this threshold, $\re \bar Y_1(\omega)$ starts to grow linearly, $\re \bar Y_1(\omega) \propto \omega-2\Delta$. The $\re \bar Y_2 (\omega) $ term, corresponding to the creation of one quasiparticle in the Andreev level and one in the continuum, is given by:
\begin{widetext}
\begin{equation}
\frac{\re \bar Y_2^{(0)}(\omega)}{G} = \pi  \!\!\!\! \int\limits_{\Delta|\cos\frac{\phi}{2}|}^\Delta \!\!\!\! d \omega_1 \theta(\omega-\omega_1-\Delta)\frac{\omega \omega_1 - \Delta^2 (1+\cos\phi)}{\omega^2 \Delta |\sin\frac{\phi}{2}| ( \omega - 2\omega_1)} \frac{\sqrt{(\omega-\omega_1)^2-\Delta^2} \sqrt{\Delta^2-\omega_1^2}}{\sqrt{\omega_1^2 - \Delta^2 \cos^2\frac{\phi}{2}}} \, . \label{reYbar2}
\end{equation}
\end{widetext}
The threshold frequency of this term is $\Delta + \Delta|\cos\frac{\phi}{2}|$, same as the threshold frequency of the fully transmitting channel for this process. Behavior near threshold is given by $\re \bar Y_2 (\omega) \propto  \omega- \Delta -\Delta |\cos\frac{\phi}{2}|$. Finally, there is a term coming from the processes in which a pair of quasiparticles is created in the Andreev level. It has the threshold frequency of $\omega_{th}(\phi)$ and is given by:
\begin{widetext}
\begin{align}
&\frac{\re \bar Y_3^{(0)}(\omega)}{G} = \theta(\omega - \omega_{th}(\phi)) \theta(2\Delta - \omega) \frac{\pi^2}{|\sin\frac{\phi}{2}|}\frac{\Delta}{\omega} \left(1 - \frac{\omega^2}{4\Delta^2}\right)  \sqrt{1 - \frac{\omega_{th}^2(\phi)}{\omega^2}} , \quad \omega_{th}(\phi) = 2\Delta |\cos\frac{\phi}{2}| \,. \label{reYbar3}
\end{align}
\end{widetext}

In the presence of quasiparticles, the above expressions for the admittance acquire additional factors reflecting the quasiparticles distribution function, similar to the single-channel junction. In addition, there are two other terms, $\bar Y_4$ and $\bar Y_5$, coming from band-to-band transitions and ionization of Andreev level, respectively. These are obtained by averaging Eqs.~(\ref{reY4}) and (\ref{reY5n}) over transmission coefficients. The complete expression for the dissipative part of the admittance in the presence of quasiparticles is given in Appendix B.

\begin{figure}
\begin{center}
\includegraphics[trim = 0mm 0mm 0mm 0mm, clip,scale=0.47]{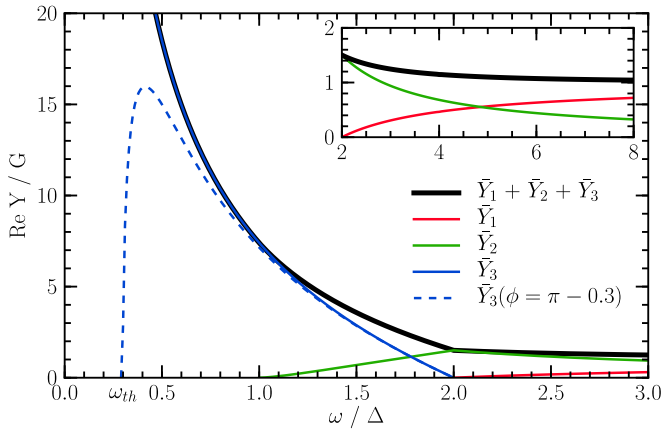}
\caption{Real part of admittance of a weak link. Solid lines
  correspond to $\phi = \pi$ case. The explicit formulas are given by Eqs.~ (\ref{reYbar1}), (\ref{reYbar2}) and (\ref{reYbar3}). $\re \bar Y_2$
  and $\re \bar Y_1$ exhibit threshold behavior at $\omega=\Delta$ and
  $\omega= 2\Delta$ where they start to grow as $\propto
  (\omega-\Delta)^{3/2}$ and $\propto \omega-2\Delta$,
  respectively. $\re \bar Y_3(\omega)$ is diverging for $\omega \to
  0$. Dashed line shows $\re \bar Y_3(\omega)$ at phase different than $\pi$, with threshold frequency $\omega_{th}$, see Eq.~(\ref{reYbar3}). As $\phi \to \pi$ its maximum grows and shifts towards the $\omega = 0$.     }\label{fig2}
\end{center}
\end{figure}

We expect the greatest change in admittance from the simple Josephson junction
at $\phi \approx \pi$, when Andreev level energies of the channels
contributing to the admittance fill the whole range of energies
between 0 and $\Delta$. In that case, for $\omega_{th}<\omega< \Delta$ and no quasiparticles present, the only
contribution to dissipative part of the admittance comes from $\bar Y_3(\omega)$ and is given by Eq.~(\ref{reYbar3}) with $\omega_{th} = \Delta |\phi - \pi|$. At this threshold $\re \bar Y_3 (\omega)$ grows as $(\omega -\omega_{th})^{1/2}$ and reaches maximum for $\omega = \sqrt{2}\omega_{th}$. The height of the maximum scales as  $\Delta/\omega_{th}$. Frequency dependence of $\re \bar Y _3$ for $\phi$ close to $\pi$  is shown on Fig. 2.

When $\phi = \pi$ exactly, there is no low-frequency cut-off. For low frequencies $\re \bar Y_3 (\omega)$ diverges as $1/\omega$. At frequencies $\Delta <\omega<2\Delta$, in addition to $\re \bar Y_3$, there is the contribution $\re \bar Y_2$ given by Eq.~(\ref{reYbar2}). Its behavior near threshold for $\phi=\pi$ is different than for any other $\phi$ and is given as $\re \bar Y_2 \propto (\omega- \Delta)^{3/2}$. At frequencies higher than $2\Delta$, the $\re \bar Y_3$ contribution vanishes and $\re \bar Y = \re \bar Y_1 + \re \bar Y_2 $. The frequency dependence of $\re \bar Y (\omega)$ for $\phi=\pi$ is also shown on Fig. 2.

If $\phi \neq \pi$, the dissipative part of the admittance is zero for $\omega< \omega_{th}$ and vanishing occupation factors. Assuming Boltzmann distribution $f(E) = A e^{-E/T}$ for quasiparticles and considering frequencies $\omega<\omega_{th}$, the only contribution to $\re \bar Y (\omega)$ comes from $\bar Y_4$ and $\bar Y_5$  terms -- Eqs.~(\ref{reYbar4}) and (\ref{reYbar5}). The former comes from transitions within the continuum band. In the limit $\omega, T \ll \Delta(1 -  |\cos\frac{\phi}{2}|)$ the most important are transitions from the bottom of the band and we get:
\begin{align}
\frac{\re \bar Y_4 (\omega)}{G} &\approx \frac{x_{qp}}{\sqrt{2\pi}} \cot^2\frac{\phi}{2} (1-e^{-\omega/T}) \sqrt{\frac{\Delta}{T}} U(\frac{\omega}{T}) \, , \label{reYbar4as}
\end{align}
where $U(x) = \int_0^\infty dt e^{-xt} \sqrt{t(1+t)}\ln(1+1/t)$. The $\re \bar Y_5 (\omega)$ term is due to transitions from Andreev levels to the continuum. In the same limit of small frequency and temperature, it is given by:
\begin{align}
\frac{\re \bar Y_5 (\omega)}{G} &\approx \frac{\pi^2 x_{qp}}{\sqrt{2\pi}} \cot^2\frac{\phi}{2} \frac{\sqrt{\Delta T}}{\omega} \sinh{\frac{\omega}{2T}}  I _1 (\frac{\omega}{2T}) \, , \label{reYbar5as}
\end{align}
where $I_1(x)$ is the modified Bessel function of the first kind. The dissipative part of admittance is given by the sum of the two terms: $\re \bar Y (\omega) = \re \bar Y_4(\omega) + \re \bar Y_5 (\omega)$. For $\omega \ll T$ the leading term comes from Eq.~(\ref{reYbar4as}) and is
frequency-independent, $\re\bar Y_4 \propto
\sqrt{\Delta/T}$. Comparing the considered case of a weak link to a
tunnel junction of the same conductance $G$, $\re \bar Y_4 (\omega)$
has an additional, small factor of $T/(\Delta
\sin^2\frac{\phi}{2}\ln(T/\omega))$, suggesting that $\re \bar
Y_4(\omega)$ is reduced in the case of a weak link at low frequencies. In the opposite limit, $\omega \gg T$, the leading term comes from Eq.~(\ref{reYbar5as}) due to higher population of low energy Andreev levels. In that case, $\re \bar Y_5(\omega) \propto x_{qp}e^{\omega/T} \sqrt{\Delta}T \omega^{-3/2}$. Because of the large exponential factor, the dissipation in the weak link is greater than in the tunnel junction of the same conductance at high frequencies.

\section{Fluctuations of admittance}
\label{fluctuation}
Superconducting junctions are crucial elements of superconducting qubits. The admittance of a junction affects the properties of such qubits (i.e. their frequency and relaxation rates)~\cite{catelani1}. As shown above, the admittance depends on the number of quasiparticles in the junction. Fluctuations of the occupation numbers cause fluctuation of the admittance. Consequently, the resonant frequency of a qubit containing the junction will fluctuate.

The inductive $1/L_J$ term (which determines the frequency of the qubit) depends only on the occupation numbers of Andreev level. Therefore, its variance depends only on the variance of the occupation numbers of Andreev level 
\begin{equation}
\left(L_J^{(0)} \right)^2  \var \frac{1}{L_J} =   (p_0+p_2 -(p_0-p_2)^2)  \, ,\label{varLJ1}
\end{equation}
The relative fluctuations of $I_J$ and $1/L_J$ are significant unless $p_0$, $p_2$ or $p_\uparrow+p_\downarrow$ are close to 1. Assuming  equilibrium between quasiparticles in the band and in the Andreev level, as well as low quasiparticle occupation numbers of the Andreev level, Eq.~(\ref{varLJ1}) reduces to:
\begin{align}
\left(L_J^{(0)} \right)^2 \var & \frac{1}{L_J} = \frac{x_{qp}}{\sqrt{2\pi}} (\Delta/T)^{1/2}  e^{(\Delta-E_A)/T}   \, , \label{varLJ}
\end{align}

The $Y_i$ terms in the expression (\ref{Ysum}) depend on the occupation numbers of the continuum states as well. However, the fluctuations of admittance caused by the fluctuations of these occupation numbers are inversely proportional to the volume of the system and therefore are negligible in the macroscopic limit. The expression for the variance of $Y(\omega)$ is then similar to the one for $\var 1/L_J$. At frequencies $\omega<\Delta+E_A$ and $\omega \neq 2 E_A$ to avoid the resonance, we get:
\begin{align}
\var \re Y(\omega) \approx   (p_0+p_2 -(p_0-p_2)^2)  (\re Y_5^{(0)} (\omega))^2\,,
\end{align}
where $\re Y_5^{(0)} (\omega)$ is given by (\ref{reY5}). Note that at low frequencies $\re Y_5(\omega)=0$ as it has a phase-dependent threshold.

To calculate the fluctuations in disordered weak links one must integrate the above expressions
for variances over the distribution of the transmission
coefficients. Assuming again equilibrium between band states and
Andreev levels, the fluctuations of the Josephson current are given by:
\begin{align}
\var I_J = \frac{\pi}{2}G \Delta^2 x_{qp} \frac{|\cos\frac{\phi}{2}|^{1/2} (1 \!-\! |\cos\frac{\phi}{2}|)}{|\sin\frac{\phi}{2}|} e^{(1-|\cos\frac{\phi}{2}|)\Delta/T}\,.
\end{align}
Here we also assumed $ T \ll \Delta(1- |\cos\frac{\phi}{2}|)$, so that the main contribution comes from the channels with lowest $E_A$ (fully transmittive channels), and $T \ll \Delta |\cos\frac{\phi}{2}|$, see also Eq.~(\ref{LJs}). Under the same assumptions, the variance of the mean inductance $1/\bar L_J$ is:
\begin{align}
\left( \bar L_J ^{(0)}  \right)^2 \var \frac{1}{\bar L_J} = \frac{e^2}{2 \pi G} x_{qp} e^{(1-|\cos\frac{\phi}{2}|)\Delta/T} \cdot g(\phi) \,, \label{varbarLJ} \\
g(\phi)=\frac{|\cos\frac{\phi}{2}|^{5/2} }{|\sin\frac{\phi}{2}| (1-\sin\frac{\phi}{2} \artanh\sin\frac{\phi}{2})^2}\,. \end{align}
The factor $ {e^2}/{2 \pi G}$ can be interpreted as $1/N_e$, where $N_e$ is the effective number of channels in the weak link. Comparing to the case of a weak tunneling junction with the same number of channels, the relative fluctuations of $1/\bar L_J$ have a factor $\exp({(1-|\cos\frac{\phi}{2}|)\Delta/T})$. This large exponential factor suggests that the fluctuations are greater in the weak link. Such shot-to-shot fluctuations contribute to inhomogeneous broadening and limit the usefulness of weak links in superconducting qubits.

\section{Non-linear absorption rate at resonant frequency}
\label{nonlinabs}
For frequencies $\omega \approx 2 E_A$ we found that the admittance of a single-channel junction has a resonant delta-function peak corresponding to creation of quasiparticles at the Andreev level. Using Eq.~(\ref{reY3}), we may re-cast the absorption rate Eq.~(\ref{absorptionrate}) in the form
\begin{equation}
W=\frac{\pi}{2}\Omega_R^2\delta(\omega-2E_A)\,.
\label{Y3-Rabi}
\end{equation}
Here 
\begin{align}
\Omega_R = |\phi_1|  \frac{(\Delta^2 - E_A^2) \sqrt{E_A^2 - \Delta^2 \cos^2 \frac{\phi}{2}}}{\Delta E_A |\sin \frac{\phi}{2}|} \,
\label{OmegaR}
\end{align}
has the meaning~\cite{bergeret2} of Rabi frequency for the transitions in an effective two-level system driven by AC perturbation $\phi_1$. The two levels correspond, respectively, to the empty and doubly-occupied Andreev state. In the linear response, we neglect the effect of Rabi oscillations on the dynamics of the two-level system. This is possible as long as $\Omega_R$ is smaller than some ``natural'', independent of $\phi_1$ width $\eta_0$ of the levels. Such natural width coming, {\sl e.g.}, from inelastic scattering of quasiparticles~\cite{rodero} leads to a replacement $\delta(\omega-2E_A)\to\eta_0/\pi[(\omega-2E_A)^2+\eta_0^2]$.

The effect of a stronger AC perturbation is two-fold. First, it may result in $\Omega_R>\eta_0$ affecting the dynamics of the two-level system. Second, it may make the levels lifetimes dependent on $\phi_1$ by introducing new processes in the kinetics of quasiparticles. Indeed, the AC field may ``ionize'' the Andreev state, transferring a quasiparticle from that state into the continuum. One needs $\omega>\Delta-E_A$ for that. Using the resonance condition, $\omega\approx 2E_A$, we find that the kinetics of the Andreev states is sensitive to the AC perturbation at $E_A>\Delta/3$.

To address these two effects, we truncate the time-dependent part of the original Hamiltonian~(\ref{calH}) retaining only terms responsible for the Rabi oscillations between the empty and doubly-occupied Andreev state and terms causing the ionization of that state,
\begin{align}
H = &\sum_\sigma E_A \alpha_\sigma ^\dagger \alpha_\sigma + \sum_{k\sigma} E_k \alpha_{k\sigma}^\dagger \alpha_{k\sigma} + \Omega_R \cos\omega t [ \alpha_\uparrow \alpha_\downarrow \notag\\
&+ \alpha_\downarrow ^\dagger \alpha_\uparrow^\dagger  ] + \cos\omega t  \sum_{k\sigma}[ \lambda_k \alpha_{k\sigma} ^\dagger \alpha_\sigma + \lambda_k^* \alpha_\sigma^\dagger \alpha_{k\sigma}] \,. \label{Hn}
\end{align}
Here $\alpha_{\sigma}$ and $\alpha_{k\sigma}$ are annihilation operators of quasiparticles in the Andreev level and the band, respectively. The last sum in Hamiltonian~(\ref{Hn}) is responsible for the transitions between the Andreev state and continuum. The corresponding ionization rate is
\begin{align}
\eta=\frac{\pi}{2}\sum_k & |\lambda_k|^2 \delta (\omega - E_k+E_A) \,.
\end{align}
The very same term leads to the $\re Y_5$ part of admittance in the linear response theory, allowing us to relate $\eta$ to $\re Y_5$, 
\begin{align}
\eta=\frac{\phi_1^2}{e^2}E_A \re Y_5^{(0)}(2E_A) \,.
\label{eta}
\end{align}

The Hamiltonian (\ref{Hn}) is quadratic, so the equations of motion for operators $\alpha$ reduce to a linear system of differential equations. Assuming frequencies close to the resonance, $\omega-2E_A \ll \Delta- E_A$, we can find the behavior of the solutions to the equations of motion after a long period of time, $t\gg 1/\eta$. The system then reaches the stationary state in which the energy absorption rate $P$ is given by:
\begin{align}
P = \langle \dot H  \rangle = \frac{3}{2} n_A \omega  \eta \, , \label{P}
\end{align}
with $n_A$ being the average number of quasiparticles in the Andreev level in the process of Rabi oscillations,
\begin{align}
n_A = \frac{\Omega_R^2}{(\omega - 2E_A)^2 + \Omega_R^2 + \eta^2} \label{nA}\,.
\end{align}
Hereinafter we neglected a shift of the resonant frequency, $|2\tilde E_A-2E_A|\propto\phi_1^2$, which is parametrically smaller than the broadening due to the $\Omega_R^2+\eta^2$ term in the denominator of Eq.~(\ref{nA}). The expression for the absorption power Eq.~(\ref{P}) has a simple interpretation: $\eta$ is the transition rate from the level to the band, so $n_A \eta$ is the rate at which the Andreev level loses quasiparticles. To keep the number of quasiparticles in the level stationary, for each particle that left, a new one must be created in the level. This amounts to energy of $\omega+E_A = 3\omega/2$ for each transition, explaining the factor of $3/2$ in Eq.~(\ref{P}). The condition $E_A>\Delta/3$ needed for $\eta\neq 0$ and the resonant condition $\omega\approx 2E_A$ imply $\omega>2\Delta/3$ for the absorption power Eq.~(\ref{P}) to be finite. 

Using Eqs.~(\ref{nA}), (\ref{eta}), (\ref{reY5}), (\ref{rho}), (\ref{z-ep-omega}), and (\ref{OmegaR}) for $n_A$, $\eta$, and $\Omega_R$ we can write $P$ in terms of the Rabi frequency,
\begin{align}
&P =  \frac{3\eta E_A \Omega_R^2}{(\omega  - 2 E_A)^2  +  \Omega_R^2 + \eta^2}\,, \notag\\
\eta=& \frac{\Omega_R^2 \sqrt{9E_A^2 - \Delta^2 } (E_A^2 +\Delta^2 \cos^2\frac{\phi}{2}) }{16 E_A \sqrt{\Delta^2 - E_A^2}(E_A^2 - \Delta^2 \cos^2 \frac{\phi}{2})}\,. \label{PRabi}
\end{align}
At generic values of static phase bias $\phi$ and transmission coefficient $\tau$, one has $\Omega_R\gg\eta$ as long as the perturbation is reasonably weak, $\phi_1\ll 2\pi$. In that case $n_A$ exhibits saturation at resonance, while $P$ grows linearly with the perturbation intensity $\propto\phi_1^2$. At a small static phase bias $\eta$ takes form
\begin{equation}
\eta=\frac{\Omega_R^2}{\Delta}\frac{\sqrt{8}}{\sqrt{\tau}(1-\tau)|\phi|^3}\,.
\label{eta-limit}
\end{equation}
It indicates that an increase of the excitation amplitude may result in a non-monotonic $n_A$ vs. $\Omega_R$ dependence and in saturation of $P$ at fairly low excitation strength, $\Omega_R\sim |\phi|^3\Delta\sqrt{\tau}(1-\tau)$.

The population of the Andreev states by quasiparticles drastically alters the critical current of the junction and its low-frequency properties due to the changes in the inductance. Using Eqs.~(\ref{LJ1}) and (\ref{nA}) we find
\begin{align}
\frac{1}{L_J} = \frac{1}{L_J^{(0)}} \cdot \frac{(\omega-2  E_A)^2 + \eta^2 }{(\omega-2  E_A)^2 + \Omega_R^2 + \eta^2}\,.
\end{align}
Therefore, $\Omega_R$ may be inferred experimentally from a measurement of the critical current~\cite{urbina} or from a two-tone experiment of the type~\cite{Devoret-Pop-Kamal}.

In the case of a disordered weak link, an AC voltage at frequencies $2\Delta|\cos\frac{\phi}{2}| < \omega< 2\Delta$ also populates Andreev levels with quasiparticles. If the applied voltage is low and the ionization processes are negligible, levels with energies within an interval $|E_A - \omega/2|\lesssim\Omega_R$ are substantially populated, cf. Eq.~(\ref{nA}). The resulting absorption power,
\begin{align}
P = &\theta(\omega -\frac{2}{3}\Delta) \frac{3 \pi^2 G}{40 e^2}|\phi_1|^3 \frac{(4\Delta^2 -\omega^2)^{3/2} \sqrt{9\omega^2 - 4\Delta^2}}{\omega^2 \Delta^2 \sin^2\frac{\phi}{2}}\notag\\
&\times[\frac{1}{2}\omega^2 +\Delta^2 \cos\phi + \Delta^2]\,, \label{Pwl}
\end{align}
scales as $|\phi_1|^3$ reflecting the growing with $|\phi_1|$ number of states involved in the absorption. As before, it is required that $\omega>2\Delta/3$ to allow the AC-field-induced ionization of the excited Andreev levels.

\section{Conclusion}
\label{conclusion}
The motivation for this study was two-fold. First, it came from the prospects~\cite{mooij} of using nanowires instead of tunnel junctions in qubits and other microwave devices~\cite{siddiqi}. Additional impetus for the study came from experiments~\cite{urbina} with nano-scale junctions, pointing to their extreme sensitivity to the presence of quasiparticles.

We obtained an analytical expression for a frequency-dependent admittance of a point contact of arbitrary transmission coefficient and arbitrary quasiparticle occupation factors. The results are valid even for non-equilibrium distribution of quasiparticles (see Section \ref{quasip}). The generalization to a short weak link (shorter than coherence length), is presented in Section \ref{weaklink}. We found that at low frequencies and temperatures, which are of interest in qubit devices, the dissipation of a point contact and a disordered weak link may indeed be lower than in a tunnel junction of a similar conductance. The lower dissipation is the result of the suppressed density of states, see Eqs. (\ref{reY4as}), (\ref{reYbar4as}) and (\ref{reYbar5as}) and the discussion following these equations. 

On the other hand, we have shown that at low temperatures the fluctuations of the admittance caused by the fluctuations of the Andreev level occupation can become large (see Section \ref{fluctuation}). At fixed number of conducting channels $N_e$, they are larger than the admittance fluctuations of a tunnel junction by a factor $\exp[(\Delta-E_A)/T]$, where $E_A$ is the energy of the lowest Andreev level, see Eqs. (\ref{varLJ}) and (\ref{varbarLJ}). In addition to that factor, already enhancing fluctuations, their amplitude scales as $N_e^{-1/2}$. Josephson junctions in the existing qubit devices have conductance $G\sim e^2/\pi$. An all-metallic link replacing such junction would have $N_e\sim 1$ leading to gigantic fluctuations. Situation is better for resonant devices designed for different applications~\cite{siddiqi} where $N_e\sim 100$, and at the same time the demand on the resonance frequency stability may be milder. 

The admittance of a single-channel junction exhibits a resonant behavior at frequencies $\omega \approx 2 E_A$ corresponding to the creation of pair of quasiparticles in the Andreev level. We studied in more detail the effect of the AC perturbation of such frequencies on a quasiparticle dynamics, see Section \ref{nonlinabs}. If $E_A<\Delta/3$, the system goes through Rabi oscillations between the empty and doubly-occupied Andreev level without dissipation. For $E_A>\Delta/3$ the AC perturbation also causes excitations of quasiparticles from the level to the band. The dissipation power is then non-zero and has resonant behavior, with the resonance width depending on the amplitude of the AC perturbation, Eq. (\ref{PRabi}). We found that the junction inductance follows the same behavior, therefore the Rabi frequency can be measured in a two-tone experiment. In the case of a disordered weak link, there is no dissipation at the AC perturbation frequencies lower than $2\Delta/3$. At higher frequencies, the dissipation power depends non-linearly on the AC perturbation intensity, see Eq. (\ref{Pwl}). Finally, it is worth noting that the population of Andreev level may depend non-monotonically on the intensity of perturbation, see Eqs. (\ref{nA}) and (\ref{eta-limit}). The population of separate Andreev levels may be studied in experiments~\cite{urbina} with break junctions.

\section*{Acknowledgements}
We thank M.H. Devoret, M. Houzet, H. Pothier, and R.J. Schoelkopf for stimulating discussions. This work was supported by DOE contract DEFG02-08ER46482 and in part by the Office of the Director of National Intelligence (ODNI), Intelligence Advanced Research Projects Activity (IARPA), via the Army Research Office W911NF-09-1-0369.

\appendix

\section{The complete expression for $\im Y(\omega)$}

From Eq.~(\ref{chi2}) we can get the complete expression
for the admittance, including the imaginary part. Since $Y(\omega)$ is
analytical in the upper half-plane, $\im Y(\omega) $ can also be obtained from the expressions for $\re Y (\omega)$ by Kramers-Kronig relations. At zero temperature, the contributions to the $\im Y (\omega)$ corresponding to the real parts from Eqs.~(\ref{reY1})-(\ref{reY3}) are given by
\begin{align}
&\frac{\im  Y_1 (\omega)}{G}  = \frac{1}{\pi \omega}\text{P.V.}\!\int_\Delta^\infty \!\!\! d\epsilon_1\int_\Delta^\infty \!\!\! d\epsilon_2 \rho(\epsilon_1) \rho(\epsilon_2) |z_{\epsilon_1,\epsilon_1 + \epsilon_2}|^2\notag\\
&\!\times [1 \!\!-\!\! f(\epsilon_1)\!\! -\!\! f(\epsilon_2)] \left[ \frac{1}{\omega \!- \! \epsilon_1 \!-\! \epsilon_2 } \!-\! \frac{1}{\omega \!+\! \epsilon_1 \!+\! \epsilon_2 } \! +\! \frac{2}{\epsilon_1 \!+\! \epsilon_2} \right]\,, \label{imY1}\\
&\frac{\im  Y_2 (\omega)}{G}  = \frac{\sqrt{\Delta^2  - E_A^2} }{\omega}\text{P.V.}\int_\Delta^\infty \!\! d\epsilon \rho(\epsilon)|z_{\epsilon,\epsilon+E_A}|^2 \notag\\
&\times  [p_0 \! +  \!  \frac{p_\uparrow \!+\! p_\downarrow}{2}\!\! -\!\! f(\epsilon)]\left[ \frac{1}{\omega \!- \! \epsilon \!-\! E_A} \!-\! \frac{1}{\omega \!+\! \epsilon \!+\! E_A } \! +\! \frac{2}{\epsilon \!+\! E_A} \right]\,, \label{imY2}  \\
&\frac{\im  Y_3 (\omega)}{G}  = \pi \frac{(\Delta^2-E_A^2)(E_A^2 - \Delta^2 \cos^2\frac{\phi}{2})}{\omega E_A^2} \notag\\
&\quad\times [p_0-p_2] \left[ \frac{1}{\omega - 2E_A} -  \frac{1}{\omega +2E_A }+\frac{1}{ E_A}\right]\,. \label{imY3}
\end{align}
These expressions are valid for any distribution of quasiparticles. The case of no quasiparticles present corresponds to $p_0=1$, $p_{\uparrow,\downarrow,2}=0$ and $f(\epsilon)=0$. The imaginary parts of the last two contributions, $Y_4(\omega)$ and $Y_5(\omega)$ are given by:
\begin{align}
&\frac{\im Y_4(\omega)}{G} = \frac{1}{\pi \omega}\text{P.V.}\! \int_\Delta^\infty d\epsilon_1\int_\Delta^\infty d\epsilon_2 \rho(\epsilon_1)\rho(\epsilon_2) |z_{\epsilon_1, \epsilon_1 - \epsilon_2}|^2 \notag\\
&\times  [f(\epsilon_2)\! -\! f(\epsilon_1)] \left[ \frac{1}{\omega \!-\! \epsilon_1 \!+\! \epsilon_2 }\! -\! \frac{1}{\omega \!+\! \epsilon_1 \!-\! \epsilon_2 } \!+\! \frac{2}{\epsilon_1 \!-\! \epsilon_2} \right]\,, \label{imY4}
\end{align}
\begin{align}
&\im Y_5(\omega)= \frac{\sqrt{\Delta^2  - E_A^2} }{\omega }\text{P.V.}\int_\Delta^\infty d\epsilon \rho(\epsilon)|z_{\epsilon, \epsilon-E_A}|^2\notag\\
&\!\times [\frac{p_\uparrow \!+\! p_\downarrow}{2} \!+\! p_2 \!-\! f(\epsilon)]\! \left[ \frac{1}{\omega \!-\! \epsilon \!+\! E_A }\! -\! \frac{1}{\omega \!+\! \epsilon \!-\! E_A } \!+\! \frac{2}{\epsilon \!-\! E_A} \right] \,. \label{imY5}
\end{align}

\section{Admittance of a weak link}
Let $p_0(\epsilon)$, $p_{\uparrow,\downarrow}(\epsilon)$ and $p_2(\epsilon)$ be the probabilities to have zero, one or two quasiparticles in the Andreev level with energy $\epsilon$. The occupation factor of the continuum state with energy $\epsilon$ is denoted by $f(\epsilon)$. The admittance of a disordered weak link for general occupation numbers is given by:
\begin{align}
\bar Y (\omega) = \sum_{i=1}^5 \bar Y_i(\omega) + \frac{i}{\omega \bar L_J}\,,
\end{align}
where the inductance term is:
\\

\begin{align}
 \frac{1}{\bar L_J} =\frac{\pi G \Delta}{|\sin\frac{\phi}{2}|} \int\limits_{\Delta |\cos\frac{\phi}{2}|}^{\Delta}\!\!\!\!d\epsilon  \frac{\Delta^2 \cos^2\frac{\phi}{2} - \epsilon^2 \sin^2\frac{\phi}{2}}{\epsilon^2 \sqrt{\epsilon^2 - \Delta ^2 \cos^2\frac{\phi}{2}}} [p_0(\epsilon) \!-\! p_2(\epsilon)]\,.
\end{align}

The real parts of the $\bar Y_i$ terms are given by:
\begin{widetext}
\begin{align}
\frac{\re \bar Y_1(\omega)}{G} \!=\!\theta(\omega \!-\! 2\Delta)\text{P.V.}\! \!\!\!\int\limits_{\Delta}^{\omega- \Delta}\!\!\!\! d \epsilon \frac{ \epsilon \omega \!-\! \Delta^2 (1 \!+\! \cos\phi)  }{ \Delta^2 \omega^2 | \sin\frac{\phi}{2}|(\omega \!-\! 2\epsilon)  } \frac{ \sqrt{\epsilon^2 \!-\! \Delta^2} \sqrt{(\omega \!-\! \epsilon)^2 \!-\! \Delta^2} }{\sqrt{\epsilon^2 / \Delta^2 \!-\! \cos^2\frac{\phi}{2}}}  \ln\! \frac{\sqrt{\epsilon^2/\Delta^2 \!-\! \cos^2\frac{\phi}{2}} \!+\! |\sin \frac{\phi}{2} |}{ \sqrt{\epsilon^2/\Delta^2 \!-\! \cos^2\frac{\phi}{2}} \!-\! |\sin \frac{\phi}{2}| }[1\!\!-\!\! f(\epsilon) \!\!-\!\! f(\omega \!\!-\!\! \epsilon)]\,, \label{reYbar1s}
\end{align}
\begin{align}
&\frac{\re \bar Y_2(\omega)}{G} = \pi \!\!\!\int\limits_{\Delta|\cos\frac{\phi}{2}|}^\Delta \!\!\! d \epsilon \theta(\omega-\epsilon-\Delta) \frac{\omega \epsilon - \Delta^2 (1+\cos\phi)}{\omega^2 \Delta |\sin\frac{\phi}{2}| (  \omega \!-\!  2\epsilon)} \frac{\sqrt{(\omega \!-\! \epsilon)^2 \!-\! \Delta^2} \sqrt{\Delta^2 \!-\! \epsilon^2}}{\sqrt{\epsilon^2 \!-\! \Delta^2 \cos^2\frac{\phi}{2}}} [p_0(\epsilon) \!+\! \frac{p_\uparrow(\epsilon) \!+\! p_\downarrow(\epsilon)}{2} \!-\! f(\omega \!-\! \epsilon)]\,,
 \label{reYbar2s}
\end{align}
\begin{align}
&\frac{\re \bar Y_3(\omega)}{G} = \theta(\omega - 2\Delta |\cos\frac{\phi}{2}|) \theta(2\Delta - \omega) \frac{\pi^2}{|\sin\frac{\phi}{2}|}\left( \frac{\Delta^2}{\omega^2} - \frac{1}{4}\right) \sqrt{\frac{\omega^2}{\Delta^2} - 4 \cos^2\frac{\phi}{2}} [p_0(\omega/2) - p_2(\omega/2)]\,, \label{reYbar3s}
\end{align}
\begin{align}
&\frac{\re \bar Y_4(\omega)}{G} = \int_\Delta^\infty  d \epsilon \theta(\epsilon + \omega - \Delta)[f(\epsilon) - f(\epsilon+\omega)] \frac{\sqrt{\epsilon^2 -\Delta^2} \sqrt{(\epsilon+\omega)^2 - \Delta^2}}{\omega^2 \Delta^2 |\sin\frac{\phi}{2}| (2\epsilon + \omega)}  \notag\\
&\times\Bigg\{  \frac{\omega(\epsilon\!+\!\omega) \!-\! \Delta^2(1 \!+\! \cos\phi)}{\sqrt{(\epsilon\!+\!\omega)^2/\Delta^2 \!-\! \cos^2\frac{\phi}{2}}}  \ln \frac{\sqrt{(\epsilon\!+\!\omega)^2/\Delta^2 \!-\! \cos^2\frac{\phi}{2}} \!+\! |\sin \frac{\phi}{2}|}{ \sqrt{(\epsilon \!+\! \omega)^2/\Delta^2  \!-\! \cos^2\frac{\phi}{2}} \!-\! |\sin \frac{\phi}{2}| } + \frac{\omega \epsilon \!+\! \Delta^2(1\!+\!\cos\phi)}{\sqrt{\epsilon^2/\Delta^2 \!-\! \cos^2\frac{\phi}{2}}}   \ln\! \frac{\sqrt{\epsilon^2/\Delta^2 \!-\! \cos^2\frac{\phi}{2}} \!+\! |\sin \frac{\phi}{2}|}{ \sqrt{\epsilon^2/\Delta^2 \!-\! \cos^2\frac{\phi}{2}} \!-\! |\sin \frac{\phi}{2}| } \Bigg\}\,, \label{reYbar4} 
\end{align}
\begin{align}
&\frac{\re \bar Y_5(\omega)}{G} = \pi \!\!\!\!\! \int\limits_{\Delta|\cos\frac{\phi}{2}|}^\Delta  \!\!\!\!\!\! d\epsilon \theta(\epsilon+\omega - \Delta) \frac{\omega \epsilon + \Delta^2(1+\cos\phi)}{\omega^2 \Delta |\sin\frac{\phi}{2}| (2\epsilon + \omega)}  \frac{\sqrt{(\omega \!+\! \epsilon)^2 \!-\! \Delta^2} \sqrt{\Delta^2 \!-\! \epsilon^2}}{\sqrt{\epsilon^2 - \Delta^2 \cos^2\frac{\phi}{2}}} \left[ \frac{p_\uparrow (\epsilon) \!+\! p_\downarrow(\epsilon)}{2} \!+\! p_2(\epsilon) \!-\!  f(\omega \!+\! \epsilon) \right] \,. \label{reYbar5}
\end{align}
\end{widetext}
From these we can also find imaginary parts using Kramers-Kronig relations. 

Assuming Boltzmann distribution $f(\epsilon) = A e^{-\epsilon/T}$ both below and
above the gap, for frequencies less than $2\Delta |\cos(\phi/2)|$ the only contribution to $\re Y (\omega)$ comes from (\ref{reYbar4}) and (\ref{reYbar5}). In the limit $\omega, T \ll \Delta$ we get:
\begin{align*}
\re \bar Y_4 (\omega) &\approx  \frac{G}{\sqrt{2\pi}}x_{qp}  \cot^2\frac{\phi}{2} \sqrt{\frac{\Delta}{T}}(1 \!-\! e^{-\omega/T}) U(\frac{\omega}{T})\,, \\
\re \bar Y_5 (\omega) &\approx \frac{\pi^2 G}{\sqrt{2\pi}} x_{qp} \cot^2\frac{\phi}{2} \frac{\sqrt{T\Delta}}{\omega}\sinh{\frac{\omega}{2T}}  I _1 (\frac{\omega}{2T})\,,
\end{align*}
where $U(x) = \int_0^\infty dt e^{-xt} \sqrt{x(1+x)}\ln(1+1/x)$ and
$I_1(x)$ is the modified Bessel function of the first kind. Now consider two opposite limits, $\omega \ll T$:
\begin{align}
\re \bar Y_4 (\omega) &\approx \frac{G}{\sqrt{2\pi}} x_{qp} \cot^2\frac{\phi}{2}\sqrt{\Delta/T}\,, \\
\re \bar Y_5 (\omega) &\approx  \frac{\pi^2G}{8\sqrt{2\pi}}x_{qp}  \cot^2\frac{\phi}{2} \frac{\omega}{\sqrt{T\Delta}}\,,
\end{align}
and $\omega \gg T$:
\begin{align}
\re \bar Y_4 (\omega) &\approx \frac{G}{2\sqrt{2}}x_{qp} \cot^2\frac{\phi}{2}   \frac{T\sqrt{\Delta}}{\omega^{3/2}} \ln \frac{\omega}{T}\,,\\
\re \bar Y_5 (\omega) &\approx  \frac{\pi G}{2\sqrt{2}} x_{qp} \cot^2\frac{\phi}{2}  \frac{\sqrt{T}\Delta}{\omega^{3/2}}e^{\omega/T}\,.
\end{align}

\bibliography{library}

\end{document}